\documentclass[final]{aipproc}
\layoutstyle{6x9}

\begin{document}

\newcommand{\Fig}[1]{Fig.~\ref{#1}}

\newcommand{\DAFNE}{DA$\Phi$NE}
\newcommand{\eg}{{\em e.g.}}

\newcommand{\cm}{\ensuremath{\mathrm{cm}}}
\newcommand{\Deg}{\ensuremath{^\circ}}
\newcommand{\GeV}{\ensuremath{\mathrm{GeV}}}
\newcommand{\Lpb}{\ensuremath{\mathrm{pb}^{-1}}}
\newcommand{\Lfb}{\ensuremath{\mathrm{fb}^{-1}}}
\newcommand{\Lcms}{\ensuremath{\mathrm{cm}^{-2}\,\mathrm{s}^{-1}}}
\newcommand{\MeV}{\ensuremath{\mathrm{MeV}}}
\newcommand{\m}{\ensuremath{\mathrm{m}}}
\newcommand{\mm}{\ensuremath{\mathrm{mm}}}
\newcommand{\mrad}{\ensuremath{\mathrm{mrad}}}
\newcommand{\ps}{\ensuremath{\mathrm{ps}}}
\newcommand{\Tesla}{\ensuremath{\mathrm{T}}}
\newcommand{\ub}{\ensuremath{\mu\mathrm{b}}}

\newcommand{\geqsim}{\, \raisebox{-0.6ex}{\ensuremath{\stackrel{\textstyle{>}}{\sim}}}\, }

\newcommand{\BR}[1]{\ensuremath{\mathrm{BR}(#1)}}
\newcommand{\SN}[2]{\ensuremath{#1\cdot10^{#2}}}
\newcommand{\VS}[2]{\ensuremath{#1\pm#2}}
\newcommand{\VSS}[3]{\ensuremath{#1\pm#2\pm#3}}
\newcommand{\VA}[3]{\ifthenelse{\equal{#2}{#3}}
{\ensuremath{#1\pm#2}}{\ensuremath{#1\,^{+#2}_{-#3}}}}
\newcommand{\Width}[1]{\ensuremath{\Gamma(#1)}}

\newcommand{\amu}{\ensuremath{a_\mu}}
\newcommand{\amuHadr}{\ensuremath{a_\mu^\mathrm{had}}}
\newcommand{\Reeps}{\ensuremath{\mathrm{Re}\,(\epsilon'/\epsilon)}}
\newcommand{\Rexp}{\ensuremath{\mathrm{Re}\ x_+}}
\newcommand{\phiP}{\ensuremath{\varphi_P}}
\newcommand{\spi}{\ensuremath{s_\pi}}
\newcommand{\Vud}{\ensuremath{|V_{ud}|}}
\newcommand{\Vus}{\ensuremath{|V_{us}|}}
\newcommand{\Vusfp}{\ensuremath{|V_{us}|f_+^{K^0\pi^-}(0)}}

\newcommand{\q}{\ensuremath{q}}
\newcommand{\qbar}{\ensuremath{\bar{q}}}
\newcommand{\qg}{\ensuremath{g}}

\newcommand{\Pazero}{\ensuremath{a_0}}
\newcommand{\Pe}{\ensuremath{e}}
\newcommand{\Pem}{\ensuremath{e^-}}
\newcommand{\Pep}{\ensuremath{e^+}}
\newcommand{\Peta}{\ensuremath{\eta}}
\newcommand{\Petapr}{\ensuremath{\eta'}}
\newcommand{\Pfzero}{\ensuremath{f_0}}
\newcommand{\Pg}{\ensuremath{\gamma}}
\newcommand{\PK}{\ensuremath{K}}
\newcommand{\PKm}{\ensuremath{K^-}}
\newcommand{\PKp}{\ensuremath{K^+}}
\newcommand{\PKpm}{\ensuremath{K^\pm}}
\newcommand{\PKS}{\ensuremath{K_S}}
\newcommand{\PKL}{\ensuremath{K_L}}
\newcommand{\Pmu}{\ensuremath{\mu}}
\newcommand{\Pnu}{\ensuremath{\nu}}
\newcommand{\Pnubar}{\ensuremath{\bar{\nu}}}
\newcommand{\Pphi}{\ensuremath{\phi}}
\newcommand{\Ppi}{\ensuremath{\pi}}
\newcommand{\Ppim}{\ensuremath{\pi^-}}
\newcommand{\Ppin}{\ensuremath{\pi^0}}
\newcommand{\Ppip}{\ensuremath{\pi^+}}
\newcommand{\Prho}{\ensuremath{\rho}}
\newcommand{\Prhop}{\ensuremath{\rho^+}}
\newcommand{\Prhom}{\ensuremath{\rho^-}}
\newcommand{\Prhon}{\ensuremath{\rho^0}}
\newcommand{\Ptau}{\ensuremath{\tau}}

\newcommand{\DKeIII}{\ensuremath{K_{e3}}}
\newcommand{\DKlIII}{\ensuremath{K_{\ell3}}}
\newcommand{\DKeIIIp}{\ensuremath{K^+_{e3}}}
\newcommand{\DKeIIIn}{\ensuremath{K^0_{e3}}}
\newcommand{\DKmuIII}{\ensuremath{K_{\mu 3}}}
\newcommand{\DKpipi}{\ensuremath{K\rightarrow\pi\pi}}

\newcommand{\DKLgg}{\ensuremath{K_L\rightarrow\gamma\gamma}}
\newcommand{\DKLmupmum}{\ensuremath{K_L\rightarrow\mu^+\mu^-}}
\newcommand{\DKLpipi}{\ensuremath{K_L\rightarrow\pi\pi}}
\newcommand{\DKLpippim}{\ensuremath{K_L\rightarrow\pi^+\pi^-}}
\newcommand{\DKLpippimpin}{\ensuremath{K_L\rightarrow\pi^+\pi^-\pi^0}}
\newcommand{\DKLpinpin}{\ensuremath{K_L\rightarrow\pi^0\pi^0}}
\newcommand{\DKLpinpinpin}{\ensuremath{K_L\rightarrow3\pi^0}}

\newcommand{\DKSpinpin}{\ensuremath{K_S\rightarrow\pi^0\pi^0}}
\newcommand{\DKSpipi}{\ensuremath{K_S\rightarrow\pi\pi}}
\newcommand{\DKSpippim}{\ensuremath{K_S\rightarrow\pi^+\pi^-}}
\newcommand{\DKSpippimg}{\ensuremath{K_S\rightarrow\pi^+\pi^-\gamma}}
\newcommand{\DKSpippimrad}{\ensuremath{K_S\rightarrow\pi^+\pi^-(\gamma)}}

\newcommand{\DKpmeIII}{\ensuremath{K^\pm_{e3}}}
\newcommand{\DKpmpipmpinpin}{\ensuremath{K^\pm\rightarrow\pi^\pm\pi^0\pi^0}}

\newcommand{\Dphietaprg}{\ensuremath{\phi\rightarrow\eta'\gamma}}
\newcommand{\Dphietag}{\ensuremath{\phi\rightarrow\eta\gamma}}
\newcommand{\Dphietaping}{\ensuremath{\phi\rightarrow\eta\pi^0\gamma}}
\newcommand{\Dphipinping}{\ensuremath{\phi\rightarrow\pi^0\pi^0\gamma}}
\newcommand{\Dphipippimpin}{\ensuremath{\phi\rightarrow\pi^+\pi^-\pi^0}}

\newcommand{\EEHadr}{\ensuremath{e^+e^-\rightarrow\mathrm{hadrons}}}
\newcommand{\EEpippim}{\ensuremath{e^+e^-\rightarrow\pi^+\pi^-}}
\newcommand{\EEpippimg}{\ensuremath{e^+e^-\rightarrow\pi^+\pi^-\gamma}}

\title{Recent results from KLOE at \DAFNE}

\author{
The KLOE collaboration\footnote{
The KLOE collaboration: A.~Aloisio, F.~Ambrosino, A.~Antonelli, 
M.~Antonelli, C.~Bacci, G.~Bencivenni, S.~Bertolucci, C.~Bini, 
C.~Bloise, V.~Bocci
% ...
}\\ 
Presented by Matthew Moulson}
{address={Laboratori Nazionali di Frascati, 00044 Frascati RM, Italy}}

\begin{abstract}
The KLOE experiment at \DAFNE\ collected about $450~\Lpb$ of data in 
2001--2002. Much of this data set has been analyzed and has yielded 
definitive results on \PKS\ and radiative \Pphi\ decays, 
as well as studies concerning a wide range of topics in 
kaon and hadronic physics.
\end{abstract}

\maketitle

KLOE is a large, general-purpose detector with optimizations for the 
study of discrete symmetries in the neutral kaon system. The experiment 
is permanently installed at \DAFNE, the Frascati \Pphi\ factory, an 
\Pep\Pem\ machine with $W\approx m_\phi \approx 1.02~\GeV$. 
The \DAFNE\ design luminosity is \SN{5.3}{32}~\Lcms. 
\Pphi's are produced with a cross section of $\sim3.2~\ub$,
and decay into \PKp\PKm\ and \PKS\PKL\ pairs with 
branching ratios (BR's) of $\sim49\%$ and $\sim34\%$.
These pairs are produced 
in a pure $J^{PC} = 1^{--}$ quantum state, so observation of a \PKS\
in an event signals the presence of a \PKL\ and vice versa. 
With an appropriate tagging technique, highly pure and nearly monochromatic 
\PKS, \PKL, \PKp, or \PKm\ beams can be obtained.

The KLOE detector consists essentially of a large drift chamber surrounded by 
an electromagnetic calorimeter.
The drift chamber~\cite{KLOE:DC} is 4~\m\ in diameter and 3.3~\m\ in length, 
which results in a fiducial volume for \PKL\ decays 
that extends to about half of a decay length. 
The momentum resolution for tracks with $p\geqsim 100~\MeV$ and 
$\theta>45\Deg$ is $\sigma_p/p \leq 0.4\%$.
The lead/scintillating-fiber calorimeter~\cite{KLOE:EmC}
consists of a barrel and two endcaps and covers 98\% of the solid angle. 
The energy resolution is $\sigma_E/E = 5.7\%/\sqrt{E (\GeV)}$. 
The intrinsic timing resolution is 
$\sigma_t = 54~\ps/\sqrt{E (\GeV)} \oplus 50~\ps$, which allows photon 
vertices from \Ppin\ decays to be reconstructed with a resolution of 
$\sim1.5~\cm$.
A superconducting coil surrounding the calorimeter barrel 
provides a 0.52~\Tesla\ magnetic field.

During 2002 data taking, the maximum luminosity sustained by \DAFNE\ was 
\SN{7.5}{31}~\Lcms. Although this is lower than 
the design value, the performance of the machine during 2002 was much 
improved with respect to previous years, and the KLOE experiment was able 
to collect as much as 4.5~\Lpb\ per day. The combined KLOE 2001--2002
data set amounts to about 450~\Lpb, or 1.4 billion \Pphi\ decays.

A series of recent upgrades to the machine, including an 
overhaul of the interaction region inside the KLOE detector, is expected
to bring the design luminosity to within reach. Data taking with KLOE is 
scheduled to restart during the fall of 2003.

\paragraph{Kaon physics with KLOE}
The tagging of \PKL\ and \PKS\ decays is fundamental to all KLOE studies of 
the \PKS\PKL\ system. 
The \DKSpippim\ decay provides an efficient tag for \PKL\ decays.
\PKS's can be tagged by identifying a \PKL\ interaction in the calorimeter. 
Since the neutral kaons from \Pphi\ decays have $\beta=0.22$ at KLOE, 
the signature of such an interaction, or ``\PKL\ crash,'' is a late, 
high-energy cluster that is not associated to any track in the drift 
chamber.
In either case, reconstruction of one kaon establishes the
trajectory of the other with an angular resolution of $\sim1\Deg$ and a 
momentum resolution of $\sim2~\MeV$.

Using the \PKL\ crash to tag \PKS\ decays, KLOE has measured the ratio of 
the partial widths for the dominant \PKS\ decay modes:
$\Width{\DKSpippimrad}/\Width{\DKSpinpin} = 
\VSS{2.236}{0.003}{0.015}$~\cite{KLOE:KSpipi}.
This value was obtained using just 17~\Lpb\ of data from the 2000 run.
In addition to providing the first part of the double ratio for \Reeps, 
the measurement of this value allows determination of 
 $\chi_0 - \chi_2$, the difference in \Ppi\Ppi\ phase shifts in \DKpipi\
transitions with $I = 0$ and 2. As argued in~\cite{CDG:Kpipi}, the 
extraction of the \DKpipi\ amplitudes from the measured widths must take 
into account the effective cutoff for the detection of final state photons 
from \DKSpippimg\ decays.
Due to the tagging technique used at KLOE, the detection efficiency for such 
decays is good out to high values of the photon energy, which allows a 
fully inclusive measurement to be made. By the evaluation 
of~\cite{CDG:Kpipi}, 
the previously existing data give $\chi_0 - \chi_2 = (\VS{56}{8})\Deg$, which 
is in somewhat poor agreement with the estimates of the difference in 
strong \Ppi\Ppi\ phase shifts, $\delta_0 - \delta_2$, from chiral perturbation 
theory, $(\VS{45}{6})\Deg$~\cite{GM:Kpipi}, and from the phenomenological 
analysis of \Ppi\Ppi\ scattering data, 
$(\VS{45}{6})\Deg$~\cite{CGL:pipi}.\footnote
{The difference between $\chi_0 - \chi_2$ and $\delta_0 - \delta_2$ is that 
the former quantity includes an additional phase shift difference from 
isospin-breaking electromagnetic effects, estimated to be about 
3\Deg~\cite{CDG:Kpipi}.}
The KLOE measurement gives $\chi_0 - \chi_2 = (\VS{48}{3})\Deg$,
which considerably improves the agreement 
with the predictions from phenomenology.

KLOE has also measured the BR's for the \DKeIII\ decays of the 
\PKS. The \Ppi\ and \Pe\ assignments are made using time-of-flight
measurements, so the BR's to final states of each lepton charge 
are measured independently. Based on 170~\Lpb\ of 2001 data, KLOE obtains the 
preliminary values for the BR's to $\Ppim\Pep\Pnu$, 
\SN{(\VSS{3.46}{0.09}{0.06})}{-4}, and to
$\Ppip\Pem\Pnubar$, \SN{(\VSS{3.33}{0.08}{0.05})}{-4}.
A nonzero value for $A_S - A_L$, the difference in the semileptonic charge 
asymmetries for the \PKS\ and \PKL, would signal $CPT$ violation, either in 
the \PKS-\PKL\ mixing or in direct transitions that also violate the 
$\Delta S = \Delta Q$ rule. While $A_L$ has recently been measured with 
precision~\cite{KTeV:A_L},
these preliminary KLOE results give the first-ever measurement of 
$A_S$: \SN{(\VSS{19}{17}{6})}{-3}.  
When the semileptonic final states are not distinguished by charge, 
KLOE obtains \SN{(\VSS{6.81}{0.12}{0.10})}{-4} for the BR.
With respect to the previous KLOE measurement~\cite{KLOE:KSsemi}, this 
represents a tenfold increase in statistics accompanied by a reduction 
of the systematic error by a third.
The difference between the partial widths for the \DKeIII\ decays of the 
\PKS\ and \PKL\ can be related to \Rexp, the parameter which quantifies 
violations of the $\Delta S = \Delta Q$ rule in $CPT$-conserving transitions. 
The preliminary KLOE measurement of the charge-undifferentiated \DKeIII\ 
branching ratio of the \PKS\ gives $\Rexp = \SN{(\VSS{3.3}{5.2}{3.5})}{-3}$, 
which is comparable in significance to the CPLEAR result~\cite{CPLEAR:Rex}. 
KLOE has an additional 280~\Lpb\ of data under analysis, and forthcoming KLOE 
measurements of the \PKL\ lifetime and \DKeIII\ BR will 
allow further improvements on the value of \Rexp.

Using the \DKSpippim\ decay to tag \PKL\ decays, KLOE has recently completed 
a measurement of $\Width{\DKLgg}/\Width{\DKLpinpinpin}$. The BR for the decay 
\DKLgg\ provides interesting tests of chiral 
perturbation theory; additionally, this decay dominates the 
long-distance contribution to \DKLmupmum~\cite{DA:Kgamma}. 
For the ratio of BR's, KLOE obtains 
\SN{(\VSS{2.793}{0.022}{0.024})}{-3} from 362~\Lpb\ of 2001--2002 
data~\cite{KLOE:KLgg}. This value is comparable in significance to that
from NA48~\cite{NA48:KSgg}. The normalization sample of \DKLpinpinpin\
decays (which is downscaled by a factor of ten) provides a value for the 
\PKL\ lifetime that is comparable in statistical significance to the 
world-average value. 

KLOE is currently studying the BR's of the various \PKL\ decays 
to charged particles. As a proof of principle, an analysis of the tagged \PKL\ 
vertices in 78~\Lpb\ of 2002 data has been performed, and gives 
values for the \PKL\ BR's to \Ppip\Ppim\Ppin, \Ppi\Pmu\Pnu, 
and \Ppi\Pe\Pnu\ that are consistent with, and which have
statistical significance comparable to, the world-average values.
When a similar, dedicated analysis of \BR{\DKLpippim} is performed 
(on 429~\Lpb\ of 2001--2002 data), the value \SN{(\VS{2.04}{0.04})}{-3} 
is obtained. The systematic errors on the above values have not been fully 
evaluated yet, but are thought to be at the 1--2\% level.

In the longer term, KLOE intends to measure \Reeps\ via the double ratio:
\begin{displaymath}
1 - 6\,\Reeps = 
        \frac{\DKLpinpin}{\DKLpippim}\cdot\frac{\DKSpippim}{\DKSpinpin}.
\label{eq:double}
\end{displaymath}
The manner in which this expression is written calls attention to the fact 
that at KLOE, cancellations of experimental systematics are sought principally 
in the ratios of the BR's for the charged and neutral decay modes. 
KLOE prospects for the measurement of \Reeps\ can be summarized as follows. 
The current KLOE measurement of the ratio of \DKSpipi\ BR's has a negligible 
statistical error and a systematic error of 0.7\%. This error is expected
to be reduced to the 0.1\% level when the 2001--2002 data are analyzed, 
both because of changes to the tagging algorithm already implemented, and 
because the errors on the various corrections are determined by the 
statistics of the control samples used to obtain them. The statistical 
errors on the \DKLpipi\ BR measurements obtained using the entire
2001--2002 data set are currently at the 1.5\% level; systematic errors are 
at about the same level and work is in progress to significantly reduce them.
A measurement of the ratio of the \DKLpipi\ BR's with an overall
error at the level of a few per mil will require at least a factor of ten 
more data.

KLOE is also undertaking a comprehensive program for the study of the decays 
of the charged kaons. The most advanced analysis concerns the 
BR and Dalitz plot for the decay \DKpmpipmpinpin. Charge 
asymmetries in the rates and Dalitz plot slopes for this decay would signal
direct $CP$ violation (see \eg~\cite{MP:K3piCPV}).
Based on 188~\Lpb\ of 2001--2002 data, the KLOE value for the BR is (\VSS{1.781}{0.013}{0.016})\%~\cite{kn:taup}. 

\begin{figure}
\includegraphics[height=.23\textheight]{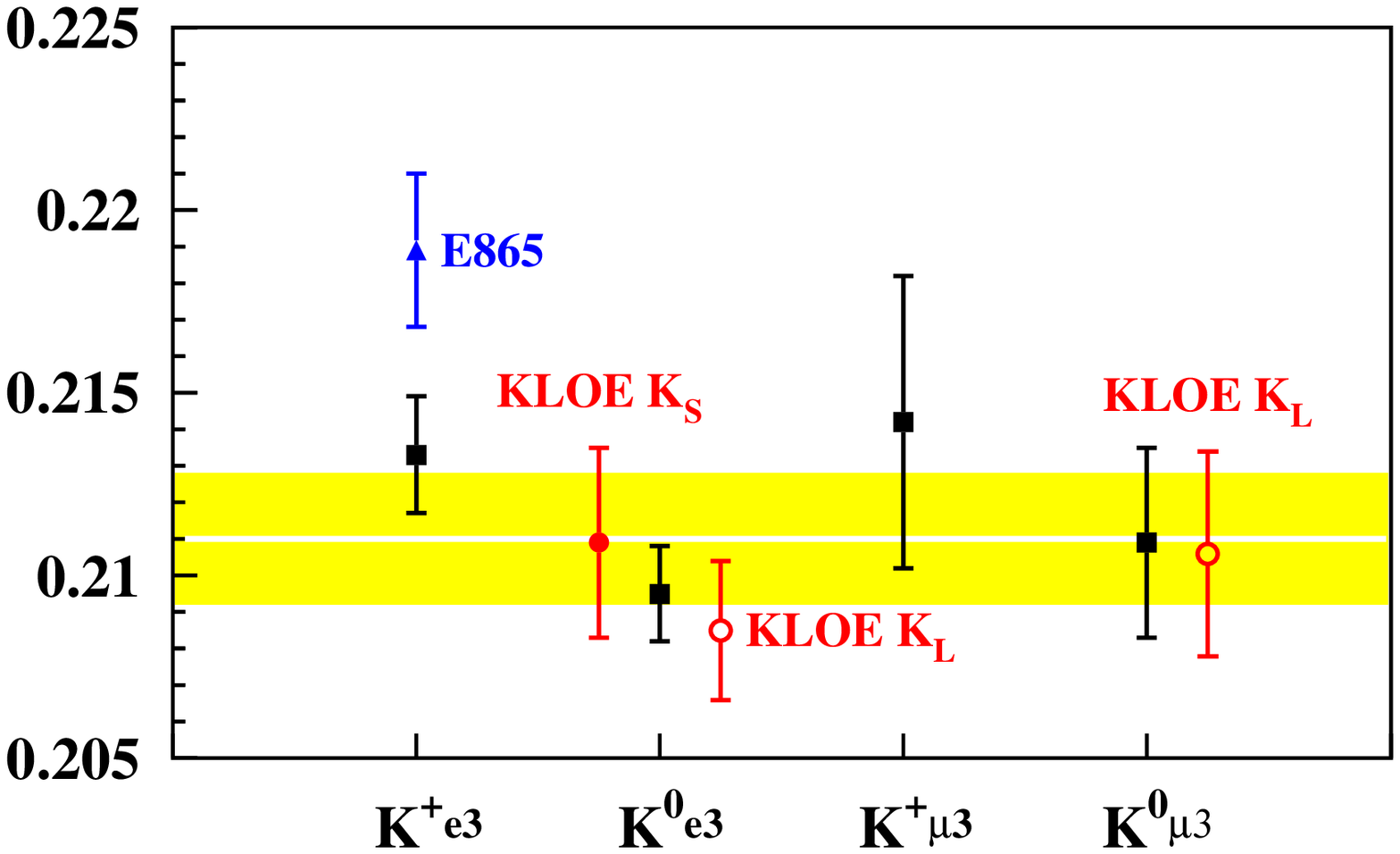}
\caption{Current status of \Vusfp. Evaluations of \Vusfp\ from the 
published world data on \DKlIII\ decays are shown as the 
squares~\cite{CKM:Vus}. The average over the \DKeIIIp\ and \DKeIIIn\ modes
(with its error) is shown as the horizontal band.
The value obtained~\cite{CKM:Vus} from the recent measurement of the \DKeIIIp\ 
BR by BNL-E865~\cite{E865:Ke3p} is shown as the triangle.
The values from the preliminary KLOE measurements of the \DKlIII\ 
decays of the \PKS\ and \PKL\ are shown as the solid and open circles.}
\label{fig:Vus}
\end{figure}

The measurement of the CKM matrix element \Vus\ represents a point of 
convergence for KLOE studies of charged and neutral kaon decays. 
\Fig{fig:Vus} summarizes values for the quantity \Vusfp\ 
obtained from measurements of the \DKlIII\ BR's and kaon 
lifetimes. The errors on these values are completely dominated by the 
experimental inputs. The four essentially independent evaluations of 
\Vusfp\ from published world data are in agreement.
The recent measurement of the \DKeIIIp\ BR by BNL-E865, 
on the other hand, gives a discrepant value. This is intriguing, because 
the value for \Vus\ obtained from the E865 measurement agrees with current 
determinations of \Vud\ given the first-row unitarity constraint on the
CKM matrix elements.
Preliminary KLOE results, on the other hand, decisively 
weigh in on the side of the existing value for \Vus. 
A KLOE measurement of the \DKeIIIp\ BR would offer direct 
comparison with the E865 result, and is forthcoming.
In the longer term, KLOE should be able to measure all four \DKlIII\ 
BR's to much better than 1\%, and to significantly improve
the determinations of the lifetimes of the \PKL\ and \PKpm, as well as 
the form factor slopes $\lambda_+$ and $\lambda_-$.

\paragraph{Hadronic physics with KLOE}
KLOE has recently completed an analysis of the decay \Dphipippimpin, 
which proceeds mainly through \Prho\Ppi\ intermediate states~\cite{KLOE:rhopi}.
A fit to the Dalitz plot containing 2 million events from 17~\Lpb\ of 2000
data gives values for the masses and widths of the \Prhop, \Prhom, and 
\Prhon.
When the fit is performed such that all three \Prho\ charge states 
are described by the same mass and width, KLOE obtains 
$m_\Prho = \VSS{775.8}{0.5}{0.3}~\MeV$ (in agreement with other recent 
\Pep\Pem\ measurements; 
see~\cite{PDG:PDB}) and 
$\Gamma_\Prho = \VSS{143.9}{1.3}{1.1}~\MeV$ (confirming the result 
of~\cite{CMD2:sighad}).
When this constraint is relaxed, no significant differences are observed 
between the masses of the charged and neutral \Prho's, or between the 
masses of the \Prhop\ and \Prhom.

The ratio of BR's for the decays \Dphietaprg\ and \Dphietag\
provides information on the \Peta-\Petapr\ mixing angle and on the gluonium 
content of the \Petapr~\cite{Fel:P,Ros:VP}. KLOE has measured this ratio 
using 17~\Lpb\ of 2000 data and obtained a value for the pseudoscalar mixing
angle in the flavor basis, $\phiP = (\VA{41.8}{1.9}{1.6})\Deg$, as well 
as a limit on the \qg\qg\ content of the \Petapr~\cite{KLOE:etapr}. 
An extension of the analysis based on the 2001-2002 data set is in progress.
 
KLOE has also analyzed the \Pphi\ decays to \Ppin\Ppin\Pg\ and 
\Peta\Ppin\Pg, where the dominant contributions involve 
production and decay of the scalar mesons \Pfzero\ and \Pazero, respectively.
Fits based on a kaon-loop model to the \Ppi\Ppi\ and \Peta\Ppi\ invariant-mass 
spectra obtained from 17~\Lpb\ of 2000 data have been used to obtain values of
the coupling constants $g_{\Pfzero\PK\PK}$ and $g_{\Pazero\PK\PK}$.
The value obtained for $g_{\Pfzero\PK\PK}$ is compatible with predictions 
that assume a \q\qbar\q\qbar\ model for the \Pfzero\ structure, while that
obtained for $g_{\Pazero\PK\PK}$ is in poor agreement with these 
predictions~\cite{KLOE:f0,KLOE:a0}. The factor-of-ten increase in
statistics from the 2001--2002 data has allowed KLOE to undertake a 
model-independent analysis of these decays featuring a complete 
study of the contributions to the Dalitz plots.

\begin{figure}
\begin{minipage}{0.79\textwidth}
\includegraphics[width=0.45\textwidth]{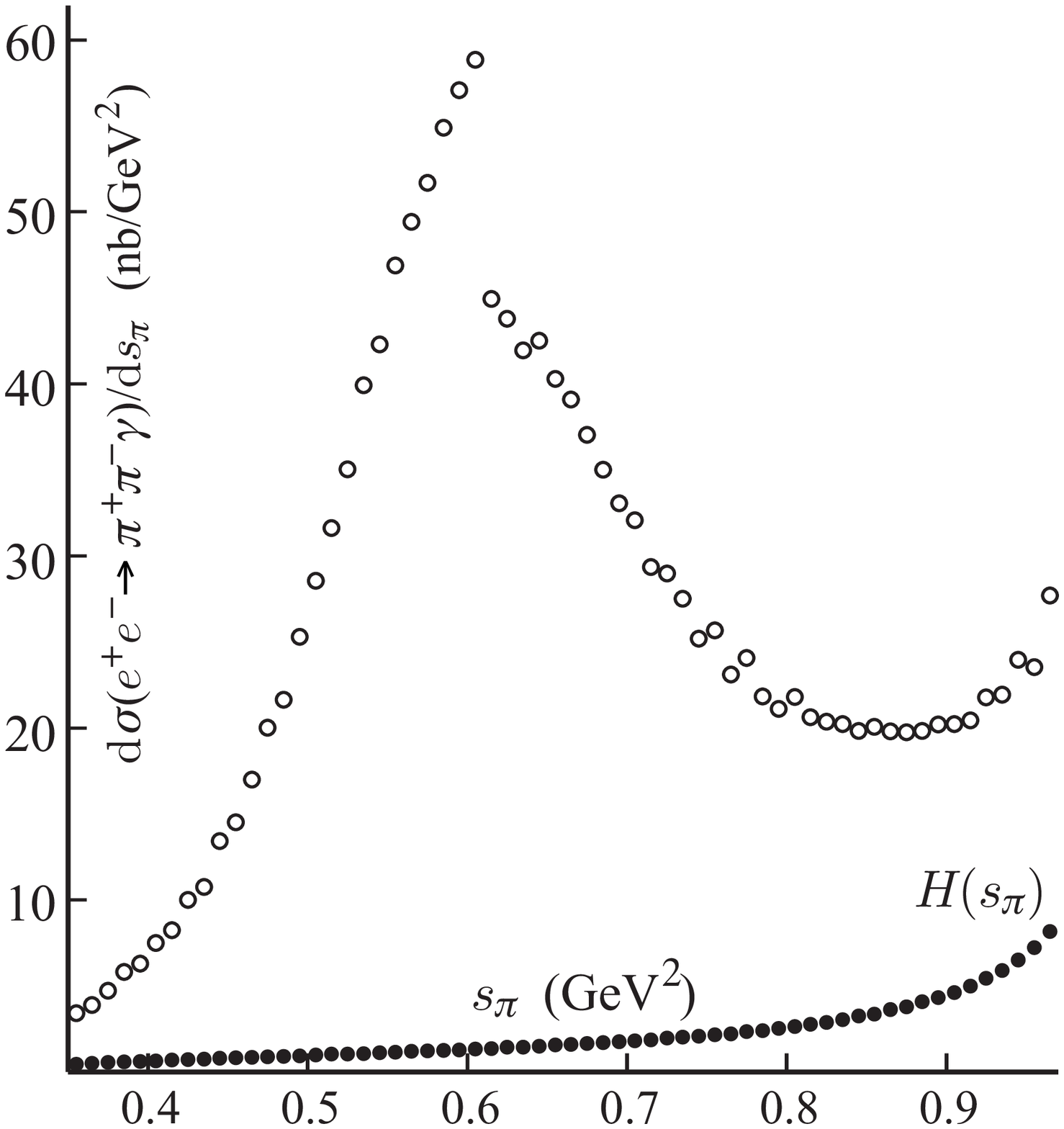}
\hspace{\fill}
\includegraphics[width=0.45\textwidth]{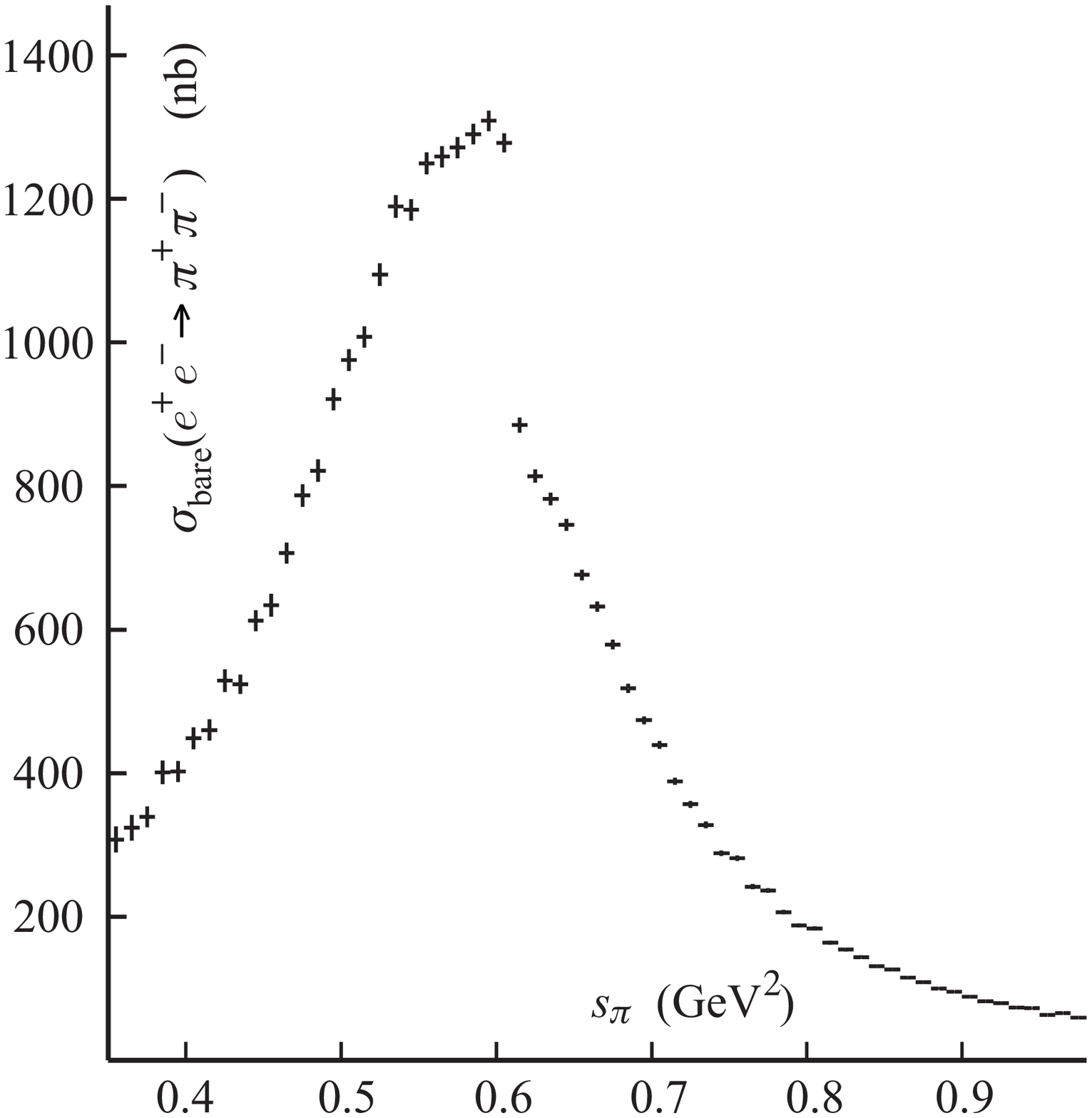}
\vspace{1.5ex}
\end{minipage}
\caption{Left: Open markers show preliminary KLOE measurements of 
$d\sigma(\EEpippimg)/d\spi$ from 1.5 million events (140~\Lpb) of 2001 data; 
solid markers show the radiation function used to extract 
$\sigma(\EEpippim)$. Right: KLOE determination of the bare cross section
for the process \EEpippim.}
\label{fig:sighad}
\end{figure}

Finally, KLOE is concluding a determination of $\sigma(\EEpippim)$
as a function of \spi, the squared CM energy of the \Ppi\Ppi\ 
system, for $0.3 < \spi < 1~\GeV^2$. \DAFNE\ operates 
exclusively at $W \approx m_\Pphi$, so the actual measurement is the 
cross section for the process \EEpippimg, where the photon is radiated 
from the initial state (ISR).
The PHOKHARA generator~\cite{KA:phok} is then used to 
relate $\sigma(\EEpippimg)$ to  $\sigma(\EEpippim)$.
Complications from processes with final-state radiation (FSR)
are avoided by restricting the current study to events
with low-angle photons ($\theta < 15\Deg$); 
in this kinematic region ISR events completely dominate the sample. 
Detection of the photon is unnecessary; \spi\ and the photon angle are 
reconstructed in the drift chamber. The preliminary KLOE data
presented in \Fig{fig:sighad} provide valuable comparison to the energy-scan 
data from CMD-2~\cite{CMD2:sighad} with respect to the determination of the 
hadronic contribution to the anomalous magnetic moment of the muon, \amuHadr.
Evaluations of \amuHadr\ based on \Pep\Pem-annihilation and \Ptau-decay data
differ by $-3.0\sigma$ and $-0.9\sigma$ from consistency
with the \amu\ measurement from BNL-E821~\cite{E821:g-2};
at present, the CMD-2 results dominate the value of \amuHadr\
from \Pep\Pem\ data~\cite{Dav:spectral}.
The KLOE data in \Fig{fig:sighad} give 
$\amuHadr(0.37 < \spi < 0.93~\GeV^2) = \SN{374.1}{-10}$, with a 
negligible statistical error and a 1.6\% systematic error including all
experimental and theoretical sources except for the lack of ISR+FSR events
in the simulation (these are already included in a new version of PHOKHARA). 
In this same interval for \spi, the CMD-2 data give
$\amuHadr = \SN{(\VSS{368.1}{2.6}{2.2})}{-10}$. The discrepancy between 
the KLOE and CMD-2 results is concentrated below the \Prho\ peak;
the values of \amuHadr\ from the two experiments agree for the 
interval $0.6 < \spi < 0.97~\GeV^2$, where the \Ppip\Ppim\ spectral 
functions for \Pep\Pem\ and \Ptau\ data disagree by 10--15\%~\cite{kn:sighad}.

\paragraph{Conclusions}
KLOE is currently analyzing a unique data set consisting of 500~\Lpb\ 
of \Pphi\ decays. Extensions of previous KLOE results on \PKS\ and 
radiative \Pphi\ decays are nearly ready, and important new measurements
such as those of \Vus\ and $\sigma(\EEHadr)$ are forthcoming. In 2003--2004
running, KLOE expects to collect a few~\Lfb\ of data, which will allow 
the broad physics reach of the experiment to be further extended.

\bibliographystyle{aipproc} 
\bibliography{moulson}

\end{document}